\documentclass[aip,jcp,amsmath,amssymb,reprint,a4paper,floatfix]{revtex4-2}

\usepackage{graphicx}
\usepackage{dcolumn}
\usepackage{bm}

\usepackage[utf8]{inputenc}
\usepackage[T1]{fontenc}
\usepackage{mathptmx}
\usepackage{etoolbox}
\usepackage{braket}
\usepackage{caption}
\usepackage{subcaption}

\makeatletter
\def\@email#1#2{%
 \endgroup
 \patchcmd{\titleblock@produce}
  {\frontmatter@RRAPformat}
  {\frontmatter@RRAPformat{\produce@RRAP{*#1\href{mailto:#2}{#2}}}\frontmatter@RRAPformat}
  {}{}
}%
\makeatother
\begin{document}

\preprint{AIP/123-QED}

\title[]{Laser-induced dynamic alignment of the HD molecule without the Born-Oppenheimer approximation}
\author{L. Adamowicz}
\affiliation{Centre for Advanced Study at the Norwegian Academy of Science and Letters, Drammensveien 78, N-0271 Oslo, Norway}
\affiliation{Department of Chemistry and Biochemistry, University of Arizona, Tucson, Arizona 85721, USA}
\author{S. Kvaal}
\affiliation{Centre for Advanced Study at the Norwegian Academy of Science and Letters, Drammensveien 78, N-0271 Oslo, Norway}
\affiliation{Hylleraas Centre for Quantum Molecular Sciences, Department of Chemistry, University of Oslo, N-0315 Oslo, Norway}
\author{C. Lasser}
\affiliation{Centre for Advanced Study at the Norwegian Academy of Science and Letters, Drammensveien 78, N-0271 Oslo, Norway}
\affiliation{Zentrum Mathematik, Technische Universit\"at M\"unchen, M\"unchen, Germany}
\author{T. B. Pedersen}
\affiliation{Centre for Advanced Study at the Norwegian Academy of Science and Letters, Drammensveien 78, N-0271 Oslo, Norway}
\affiliation{Hylleraas Centre for Quantum Molecular Sciences, Department of Chemistry, University of Oslo, N-0315 Oslo, Norway}
\email{t.b.pedersen@kjemi.uio.no}

\date{\today}
\begin{abstract}
Laser-induced molecular alignment is well understood within the framework of the Born-Oppenheimer (BO) approximation.
Without the BO approximation, however, the concept of molecular structure is lost, making alignment hard
to define precisely. In this work, we demonstrate the emergence of alignment
from the first-ever non-BO quantum dynamics simulations, using the HD molecule exposed to ultrashort laser pulses
as a few-body test case.
We extract the degree of alignment from the non-BO wave function by means of
an operator expressed in terms of pseudo-proton coordinates that mimics the BO-based definition of alignment.
The only essential approximation, in addition to the semiclassical electric-dipole approximation for the
matter-field interaction, is
the choice of time-independent explicitly correlated Gaussian basis functions.
We use a variational, electric-field-dependent basis-set construction procedure, which allows us to keep the
basis-set dimension low whilst capturing the main effects
of electric polarization on the nuclear and electronic degrees of freedom. The basis-set construction procedure is validated by
comparing with virtually exact grid-based simulations for two one-dimensional model systems: laser-driven
electron dynamics in a soft attractive Coulomb potential and nuclear rovibrational dynamics in a Morse potential.
\end{abstract}
\maketitle

\section{Introduction}

Manipulating and ultimately controlling molecules by means of electromagnetic fields~\cite{Lemeshko2013}
has the potential to fundamentally change chemistry,
as illustrated by recent work on electric-field mediated reactions and catalysis.~\cite{Stuyver2020,Shaik2020,Leonard2021}
Within spectroscopy, 
an important effect of an electric field is the alignment and possibly even orientation of
molecules with respect to the polarization direction,~\cite{Friedrich1991,Stapelfeldt2003}
paving the way for accurate investigations of the coupled electronic-nuclear dynamics down to the attosecond time scale.
While most chemical applications in condensed phases use static electric fields, the state-of-the-art for
individual molecules is impulsive laser-induced alignment, which periodically revives after the laser is turned off
(provided decoherence is avoided).~\cite{Stapelfeldt2003}

The theory of laser-induced molecular alignment~\cite{Friedrich1995,Friedrich1995a,Seideman1995}
obviously relies on the concept of molecular structure, i.e., it relies
on the Born-Oppenheimer (BO) approximation~\cite{Born1927,Born1954} in one form or another,
including the Born-Huang expansion~\cite{Born1954} which is often, incorrectly,~\cite{Sutcliffe2012,*Sutcliffe2014,Jecko2014}
claimed to be exact. In the simplest BO-based treatment, the molecule is considered a rigid rotor,
which aligns with the linear polarization direction of the laser pulse
through the generation of an angularly confined rotational wave packet. 
In exact molecular quantum mechanics, however, the concept of molecular structure is
lost due to the spherical atom-like symmetry of any molecular system combined with the particle-permutation symmetries
of the full molecular wave function.~\cite{Cafiero2004,Matyus2019,Matyus2021} In this work, we investigate if laser-induced alignment
emerges from non-BO simulations. Our goal is a qualitative theoretical confirmation of the 
overwhelming experimental evidence of the laser-induced alignment phenomenon, not quantitative
predictions of an actual experimental setup.

In past decades, explicitly correlated Gaussian (ECG)~\cite{Boys1960,*Singer1960} functions have been used
as basis functions to compute the lowest-lying (rovibrational, in BO terminology which is often used
also in non-BO work) states
with spectroscopic accuracy
without invoking the BO approximation.~\cite{Matyus2019,Bubin2013,Mitroy2013}
Although imaginary-time propagation has been investigated,~\cite{Varga2019}
the parameters of the ECGs are typically
optimized alongside the linear expansion parameters in a variational energy minimization. 
Embedded in dissociation continua, higher-lying bound rovibronic states
(corresponding to electronically excited states)
are more challenging to describe and it has been proposed to treat them as resonances.~\cite{Matyus2013}

In the present work, we investigate laser-induced alignment of the HD molecule using ECG-based simulations
without the BO approximation. Although the natural generalization of the computational approach to stationary
states would be to determine the nonlinear ECG parameters from the time-dependent variational principle, we
here adopt a simplified approach using statically optimized ECG parameters in the presence of electric fields.
Expanding the all-particle wave function in these fixed ECG basis functions leads to simple equations of motion (EOMs)
for the time-dependent linear coefficients that can be propagated using well-known integrators.
Laser-induced alignment and post-pulse revivals should then emerge clearly in the three-dimensional pseudo-proton density,
which is spherically symmetric in the initial state.

\section{Theory}

\subsection{Laser-driven non-BO quantum dynamics}

We use atomic units throughout this manuscript unless explicitly stated otherwise.
We treat the HD molecule interacting with a laser pulse in the nonrelativistic electric-dipole approximation, allowing
us to separate the internal (relative) motion from the center-of-mass motion, which is unaffected by the laser as HD is neutral.
Using the coordinate transformation of
Ref.~\onlinecite{Bubin2013} with the origin of the internal coordinate frame placed at the deuteron (with mass~\cite{Tiesinga2021}
$M_D=3670.48296785$), the internal Hamiltonian becomes $\hat{H}(t) = \hat{H}_0 + \hat{V}(t)$, where $t$ denotes time and
\begin{align}
   &\hat{H}_0 = \sum_{i=1}^3 \left( \frac{\hat{p}_i^2}{2m_i} + \frac{q_0 q_i}{r_i} \right)
             + \sum_{i=1}^2 \sum_{j=i+1}^3 \left( \frac{q_iq_j}{r_{ij}}
                + \frac{\hat{\boldsymbol{p}}_i \cdot \hat{\boldsymbol{p}}_j}{M_\text{D}}\right), \\
   &\hat{V}(t) = -\hat{D} \mathcal{E}(t).
\end{align}
The laboratory and internal frames are chosen parallel~\cite{Bubin2013}
with the $z$-axis parallel to the electric field,
\begin{equation}
\label{eq:laser}
  \mathcal{E}(t) = \mathcal{E}_0\sin(\omega (t-t_0)) G(t),
\end{equation}
where $\mathcal{E}_0$ is the field strength, $\omega$ is the carrier frequency, and $t_0 \geq 0$ is the time at which the laser is turned on.
The envelope, $G(t)$, controls the shape and
duration of the laser pulse.
The operator $\hat{D}$ is the component along the field direction of the 
electric-dipole vector operator, 
$\hat{\boldsymbol{D}} = \sum_{i=1}^3 q_i \boldsymbol{r}_i$.
The time-dependent Schr\"odinger evolution $i\,\dot\Psi(t) = \hat H(t)\Psi(t)$ is started at $t=0$ with the ground 
state of the internal Hamiltonian, $\hat H_0$. 

The Hamiltonian $\hat{H}(t)$ describes the motion of three interacting particles, referred to
as pseudo-particles,~\cite{Bubin2013} in the
central field of the charge of the deuteron,
$q_0=1$.
The position relative to the deuteron of pseudo-particle $i$ 
with charge $q_i$ ($q_i=1$ for the pseudo-proton and $q_i=-1$ for the pseudo-electrons)
and reduced mass $m_i = M_D M_{i+1} / (M_D + M_{i+1})$ 
($M_i=1$ for the pseudo-electrons and $M_i=1836.15267389$
for the pseudo-proton~\cite{Tiesinga2021})
is denoted $\boldsymbol{r}_i$, and $\hat{\boldsymbol{p}}_i$ is its conjugate momentum operator. Finally,
$r_i=\vert\boldsymbol{r}_i\vert$ and $r_{ij} = \vert \boldsymbol{r}_i - \boldsymbol{r}_j \vert$.

Using a spin-free formulation~\cite{Bubin2013,Pauncz_SGQC,*Hamermesh_GT} and 
a fixed time-independent and non-orthogonal ECG basis set of a finite dimension, $L$,
we obtain
wave functions for stationary (electronic spin-singlet) states, $\ket{\psi_n}$, and the corresponding energies, $E_n$,
by diagonalization of $\hat{H}_0$. Following Ref.~\onlinecite{Bubin2013}, the ECGs are not explicitly symmetry adapted with respect
to the rotational and inversion symmetries of $\hat{H}_0$ and, hence, the stationary states cannot be strictly
characterized by angular momentum quantum numbers and parity.
The time-dependent approximate wave function may then be expressed as
\begin{equation}
  \Psi_L(x,t) = \sum_{n=0}^{L-1} \psi_n(x) C_n(t),
\end{equation}
where the pseudo-particle coordinates are collected in the $d$-dimensional vector, $x$
($d=9$ for HD).
The expansion coefficients are determined by the time-dependent variational principle, leading to the EOMs
\begin{equation}
\label{eq:EOM}
   \text{i}\dot{C}_n(t) = E_nC_n(t) + \sum_{m=0}^{L-1} {V}_{nm}(t)C_m(t),
   \quad C_n(0) = \delta_{n0},
\end{equation}
where the dot denotes the time derivative and
the initial condition corresponds to HD being in the ground state, $\ket{\psi_0}$, before the laser is turned on at $t=t_0 \geq 0$.
The matter-field interaction matrix elements are defined as $V_{nm}(t) = \braket{\psi_n \vert \hat{V}(t) \vert \psi_m}$.

Within the nonrelativistic electric-dipole approach outlined above, the only approximation is the construction of the
finite ECG basis set, to which we turn our attention next.

\subsection{Basis construction}
The nonorthogonal basis set, $\{\ket{\phi_\mu}\}$, consists of normalized shifted ECGs~\cite{Boys1960,*Singer1960,Mitroy2013}
\begin{equation}
\phi_\mu(x) = (2^d\det(A_\mu)/\pi^d)^{1/4}\exp(-(x-s_\mu)^\dagger A_\mu (x-s_\mu)),
\end{equation}
that are defined by real symmetric positive-definite
$d \times d$ width matrices~$A_\mu$ and real shift vectors $s_\mu$ of dimension $d$.
For computational reasons, the width matrices are constrained to be spatially isotropic,~\cite{Bubin2013} but we keep the 
theoretical formulation general for notational convenience.
Depending explicitly on interparticle distances, the ECGs are excellently suited for the description of correlated motion whilst
allowing analytic evaluation of Hamiltonian integrals.~\cite{Boys1960,*Singer1960} We refer to Ref.~\onlinecite{Mitroy2013}
for a detailed review of ECGs and their applications.
We denote by $z_\mu$ the nonlinear parameter vector that contains both the entries of a Cholesky factor
$L_\mu$ of $A_\mu$ 
($A_\mu = L_\mu^T L_\mu$)
and of the shift vector, $s_\mu$. 
Referring to the width matrices by their Cholesky factors will be useful for the minimization problem we describe next.

The basis is constructed by a five-step procedure based on the minimization of the Rayleigh quotient 
\begin{equation}
R(z_1,\ldots,z_N;\mathcal E) = \frac{\sum_{\mu,\nu} c_\mu^* c_\nu 
\braket{\phi_\mu\mid \hat H_0-\hat D \mathcal E\mid \phi_\nu}}{\sum_{\mu,\nu} c_\mu^* c_\nu 
\braket{\phi_\mu\mid \phi_\nu}},
\end{equation}
for fixed values of the \emph{static} electric field $\mathcal E$.
For zero field, the minimization problem is well-defined, and one obtains an approximate ground state 
$\ket{\psi_0} = \sum_\mu \ket{\phi_\mu} c_\mu$
of the zero-field Hamiltonian, $\hat H_0$. For non-vanishing electric field, the Hamiltonian $\hat H_0-\hat D \mathcal E$ is unbounded from below.
For sufficiently small fields, however, a local minimum corresponding to a localized state is found by constraining the ranges of the nonlinear parameters.

\begin{description}
\item[First step] 
The first $N$ elements, $\ket{\phi_{0,1}},\ldots,\ket{\phi_{0,N}}$, of the basis, i.e. the nonlinear parameters $z_{0,1},\ldots,z_{0,N}$, are obtained by minimizing the Rayleigh quotient, $R(z_1,\ldots,z_N;\mathcal E = \mathcal E_0)$ with zero field $\mathcal E_0 = 0$.
\item[Second step]
An increasing grid of positive electric fields $( \mathcal E_1,\ldots,\mathcal E_M)$ is selected. For each field $\mathcal E_m$ a
set of basis functions, $\ket{\phi_{m,1}},\ldots,\ket{\phi_{m,N}}$, is obtained by minimizing $R(z_1,\ldots,z_N;\mathcal E=\mathcal E_m)$. The minimization 
is initiated by the nonlinear parameters 
determined for the previous field $\mathcal E_{m-1}$.
\item[Third step] All basis functions obtained in the second step are 
iteratively joined with the zero-field basis upon passing the following simple linear-dependence test. A basis candidate, $\ket{\phi}$, is added to
the basis if $\vert\braket{\phi\mid\phi_\mu}\vert < \tau$, where $\tau > 0$ is a chosen tolerance, for all basis functions, $\ket{\phi_\mu}$, that constitute the current basis.
\item[Fourth step] 
Due to the axial symmetry of $\hat{H}_0-\hat{D}\mathcal{E}$, the basis function $\ket{\phi_{m,\mu}}$ corresponding to the opposite field, $-\mathcal E_m$, has the same width matrix $A_{m,\mu}$ and a shift $s_{m,\mu}$ that is flipped in the opposite direction.  Such a basis function is appended to the basis if its counterpart has passed the acceptance test.
The resulting basis is called the generator set.
\item[Fifth step] Due to the spherical symmetry of $\hat{H}_0$, three copies of the generator set are obtained
by rotating the shift vectors three times consecutively about the $x$-axis by an angle of $\pi/4$. The resulting set consisting
of four copies of the generator set is then
rotated by an angle $\pi/2$ about the $y$- and $z$-axes, yielding a spherically distributed basis set consisting of nine
unique copies of the generator set.
\end{description}

\section{Results}

\subsection{Validation of the basis-set construction for 1d systems}

We test the basis-set construction procedure for two simple 
1d model systems representing a bound electron and
a bound proton 
in the HD molecule by comparing with the results from
the split-step Fourier (SSF) method,~\cite{Lubich_QCMD} which is an essentially exact grid-based approach.
For 1d systems, the fifth step of the basis-set construction procedure above is excluded.

\subsubsection{Basis for 1d Coulomb system}
We first consider the hydrogen atom in 1d.
The Hamiltonian representing the internal
motion of the system is given by
$\hat H_0 = -\frac{1}{2}
\frac{\text{d}^2}{\text{d}x^2} + V(x)$,
where, for simplicity, we have chosen unit effective mass of the electron.
The potential $V(x)$ is obtained 
from the
attractive soft Coulomb (SC) potential
\begin{equation}
V_{SC}(r) = -\frac{1}{\sqrt{r^2+\delta^2}},
\end{equation}
by setting $r = |x|$: $V(x) = V_{SC}(|x|)$.
The parameter $\delta=\sqrt{2}$ 
is chosen such that the ground state energy is 
approximately equal to $-0.5$.
The zero-field basis consists of $N=4$ origin-centered Gaussians, 
and the optimization procedure is initiated by the even-tempered ansatz,
$A_\mu = 4\cdot 1.3^{2-2\mu}$.
The grid of field values is chosen as $0.01, 0.02, 0.03, 0.032, 0.0321, 0.0322, 0.0323, 0.032394$.
No local minima are detected at higher field values.
The tolerance for basis-function acceptance is $\tau=0.98$ and the final basis consists of $L=16$ Gaussians,
which are used to diagonalize $\hat{H}_0$.

The system, initially in its ground state, is then exposed to a $100$-cycle laser pulse of the shape \eqref{eq:laser}
with envelope
\begin{equation}
\label{eq:sin2}
  G(t) = \sin^2\left( \frac{\pi (t-t_0)}{T} \right)\Theta(t-t_0)\Theta(T+t_0-t),
\end{equation}
where $\Theta(t)$ is the Heaviside step function,
$t_0 = 20$ is the time at which the laser is turned on, and $T=200\pi/\omega$ is the duration.
The carrier frequency $\omega = 0.2672$ is resonant with the lowest-lying electric-dipole-allowed
excited state and the field strength is $\mathcal{E}_0=0.01$. The propagation is done with time step $\Delta t = 0.08$. 
The reference SSF simulation, with the initial ground state obtained from inverse iteration,
is performed with the same laser pulse and time step on a uniform real-space grid of $2048$ points in the interval $[-300, 300]$.

\begin{figure}
\centering
\includegraphics[width=\columnwidth]{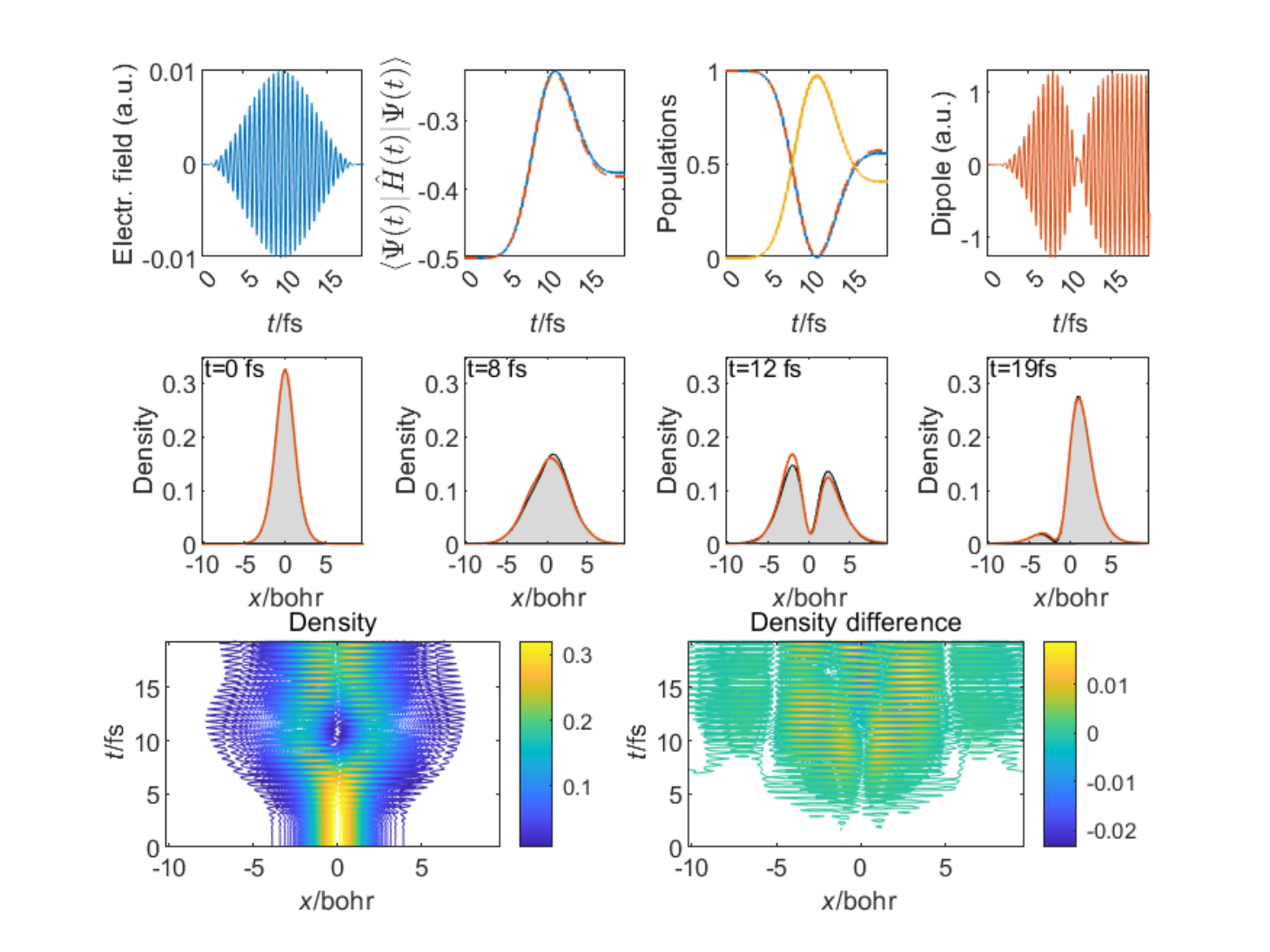}
\caption{\label{fig:1d_coulomb}Dynamics of the 1d Coulumb system.
Top row from left: $\mathcal{E}(t)$, expectation value of the Hamiltonian,
population of the ground and first dipole-allowed states, and the induced
electric dipole moment. Full and dashed lines from the Gaussian-based and SSF simulations, respectively.
Middle row: position probability densities at selected points in time. Red curves from the Gaussian-based simulation,
grey areas from the SSF simulation.
Bottom row: contour plot of the Gaussian position probability density as function of $x$ and $t$ to the left, difference between the
Gaussian and SSF densities to the right. 
}
\end{figure}

The results are shown in Fig.~\ref{fig:1d_coulomb}. The energy and induced dipole moment are very well approximated by the
Gaussian basis, whereas the position probability density shows somewhat larger errors after the peak intensity of the laser pulse. This 
is caused by ionization processes that are not captured by the Gaussians. Still, the spread and oscillations of the electronic wave packet
are largely correct with the Gaussians, showing errors of at most $10\%$. We conclude that the Gaussian basis-set construction procedure
yields roughly correct electron dynamics as long as the driving laser field does not induce significant ionization of the system. 

\subsubsection{Basis for 1d Morse system}
We consider the 1d HD molecule represented by
the Hamiltonian 
$\hat H_0 = -\frac{1}{2m}\frac{d^2}{dx^2}+V(x)$,
where the effective mass is given in terms of the
deuteron and proton masses as
$m = \frac{M_D M_p}
{M_D + M_p} = 1224$.
The potential describes the interaction
between the two nuclei, which we model
as the
H$_2$-parametrized Morse potential 
\begin{equation}
V_M(r) = D_e(1-\exp(-a(r-r_e)))^2,
\end{equation}
such that $V(x) = V_M(|x|)$.
Hence, $V(x)$ is a double-well potential with the two wells
symmetrically located at opposite sides of the $x$ axis
and separated by a large potential barrier.
The Morse parameters are $D_e=0.17449$, $a=1.4556$, and $r_e=1.4011$. 
The zero-field basis is built of $N=8$ Gaussians, half of them centered in the left well and the other half in the right well.
The grid of field values is $0.01, 0.02, 0.04, 0.06, 0.08, 0.09, 0.09354$. At higher field values, the 1d HD molecule dissociates.
The tolerance for basis-function acceptance is $\tau=0.98$ and the final basis consists of $L=26$ Gaussians,
which are used to diagonalize $\hat{H}_0$.

The system, initially in its ground state, is then exposed to the same $100$-cycle laser pulse as the 1d Coulomb system above,
except that the carrier frequency is adjusted to be resonant with the lowest-lying dipole-allowed excited state,
$\omega=0.0228$, and the laser is turned on at $t_0=275$.
The propagation is done with time step $\Delta t = 0.93$. 
The reference SSF simulation, with the initial ground state obtained from inverse iteration,
is performed with the same laser pulse and time step on a uniform real-space grid of $2048$ points in the interval $[-300, 300]$.

\begin{figure}
\centering
\includegraphics[width=\columnwidth]{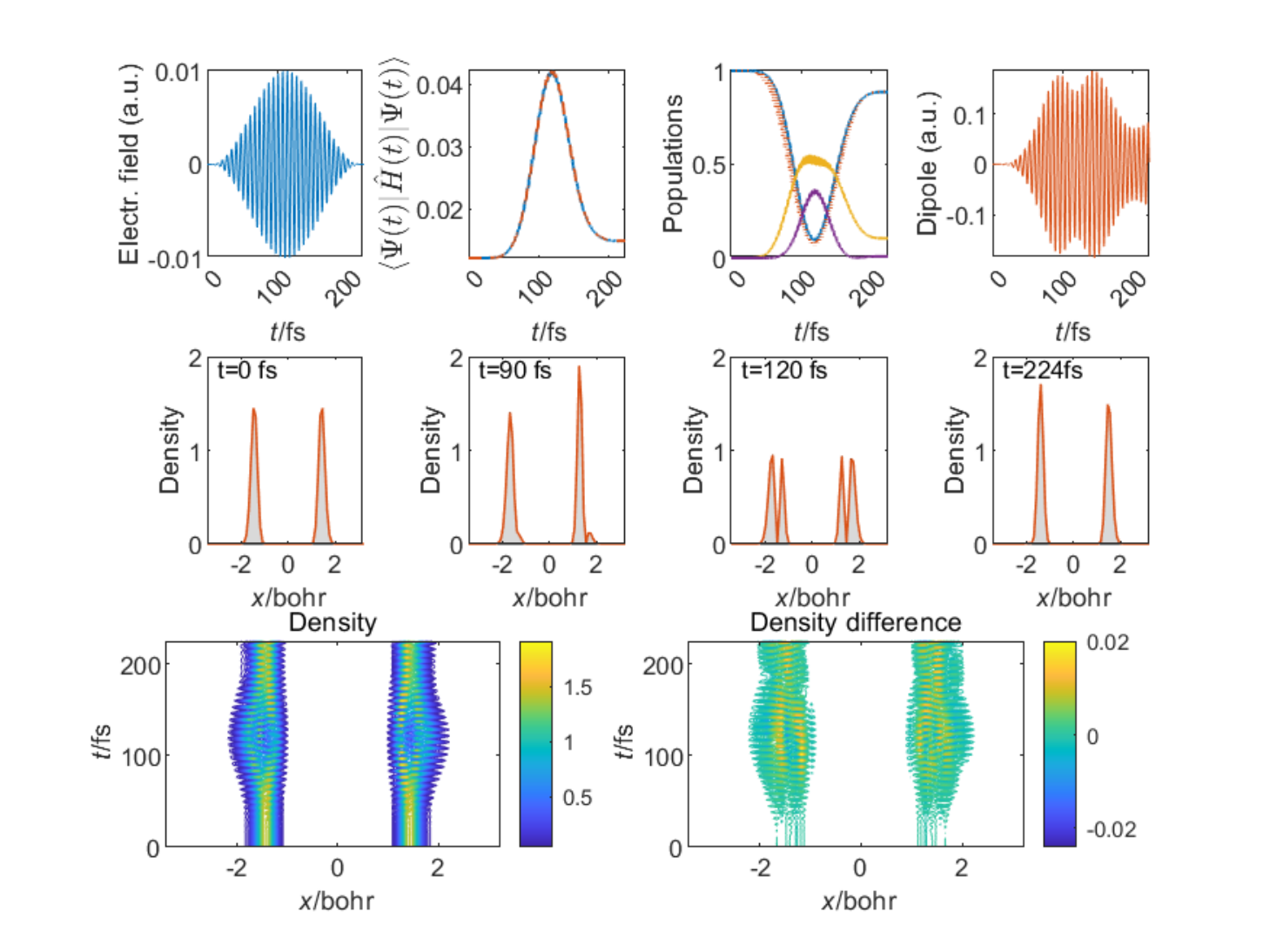}
\caption{\label{fig:1d_morse}Dynamics of the 1d Morse system.
Top row from left: $\mathcal{E}(t)$, expectation value of the Hamiltonian,
population of the ground and the two lowest-lying dipole-allowed states,
and the induced electric dipole moment. Full and dashed lines from the Gaussian-based and SSF simulations, respectively.
Middle row: position probability densities at selected points in time. Red curves from the Gaussian-based simulation,
grey areas from the SSF simulation.
Bottom row: contour plot of the Gaussian position probability density as function of $x$ and $t$ to the left, difference between the
Gaussian and SSF densities to the right. 
}
\end{figure}

As shown in Fig.~\ref{fig:1d_morse}, the agreement between the Gaussian-based and SSF simulations is excellent. The error in
the position probability density is on the order of $1\%$ at worst.
We conclude that the Gaussian basis-set construction procedure
yields roughly correct rovibrational dynamics.

\subsection{Laser-driven alignment of the HD molecule}

The basis-set construction procedure is initiated with $N=8$ ECGs at zero field. The initial values of the
nonlinear parameters are variationally
optimized within the orbital approximation (no coupling between coordinates) with centers at a distance of
$1.4$ (roughly the BO equilibrium distance of HD) from the origin on the z-axis. Relaxing the orbital approximation,
the final zero-field set is then fully optimized. 
The basis-function acceptance threshold is $\tau=0.98$ and
the grid of positive field values is chosen to be
$0.01, 0.02, 0.04, 0.06, 0.08, 0.085, 0.0913486$.
At greater fields, no stable local minima are detected.
The resulting generator set consists of $110$ ECGs and the final basis set thus consists of $L=990$ ECGs.
We stress that this is a rather modest basis-set size and that neither ionization nor full dissociation dynamics
can be properly accounted for.

The resulting ground-state energy is $-1.153798$, roughly $0.012$ above the rovibrational ground-state energy obtained within
the BO approximation using full configuration interaction (FCI) theory
with the aug-cc-pV5Z Gaussian basis set~\cite{Dunning1989,*Woon1994} to compute the potential-energy
curve. The mean HD distance is $1.497$, which yields an estimated rotational constant of $40\,\text{cm}^{-1}$
(compared with $44.6421\,\text{cm}^{-1}$ in the essentially exact vibrational ground state from the BO FCI calculation)
and a rotational period of $t_\text{rot} \approx 417\,\text{fs}$. In order to study laser-driven alignment,
we use a nonadiabatic pulse (duration shorter than $t_\text{rot}$). In agreement with BO selection rules, all rovibrational
transitions are electric-dipole forbidden and, therefore, we choose the carrier frequency resonant with the
dipole-allowed rovibronic state located $0.482681$ above the ground state, well below the experimental~\cite{Shiner1993}
ionization energy of $0.5676$. The oscillator strength of this transition is $0.14$, and the excited rovibronic state couples
to several low-lying rovibrational states with transition energies in the vicinity of the carrier frequency and with oscillator strengths
ranging from $0.02$ to $0.14$. Hence, the resonant laser pulse should populate several rovibronic states below the ionization
energy and thus induce alignment.
Although alignment induced by laser pulses resonant or near-resonant with a vibrational or robibrational transition
have been studied previously,~\cite{Seideman1995,Seideman1999,Pegarkov2000,Mainos2008}
we remark that the alignment mechanism studied in the present work does \emph{not} correspond to most experimental
setups where non-resonant carrier frequencies
are chosen to explicitly avoid electronic excitations.

The pulse is defined by Eq.~\eqref{eq:laser} with the envelope \eqref{eq:sin2} and peak intensity $3.5\,\text{TW/cm}^2$
($\mathcal{E}_0=0.01$).
The carrier frequency is $\omega=0.482681$,
the duration is $100$ cycles, $T=31.5\,\text{fs}$, and the laser is turned on at $t_0=0$.
We numerically integrate the EOMs \eqref{eq:EOM}
using the sixth-order Gauss-Legendre integrator~\cite{HairerLubichWanner_GNI,Pedersen2019} with a convergence tolerance on the
residual norm of $10^{-10}$ and time step $\Delta t = 0.1$.

To monitor alignment of the HD molecule with respect to the field direction ($z$), we compute at each time step
the expectation value of the operator
\begin{equation}
  \hat{A}_p = z_p^2 - x_p^2 - y_p^2 = r_p^2\cos(2\theta_p),
\end{equation}
where subscript $p$ indicates pseudo-proton coordinates.
In analogy with the BO-based theory,~\cite{Stapelfeldt2003} the polar angle $\theta_p$ measures the degree of alignment.
Assuming that the amplitudes of the vibrational motion are much less than $1$ throughout the dynamics, we expect maximal
alignment to occur at the maxima of $\braket{\hat{A}_p} = \braket{\Psi(t) \vert \hat{A}_p \vert \Psi(t)}$.

\begin{figure}
\centering
\includegraphics[width=\columnwidth]{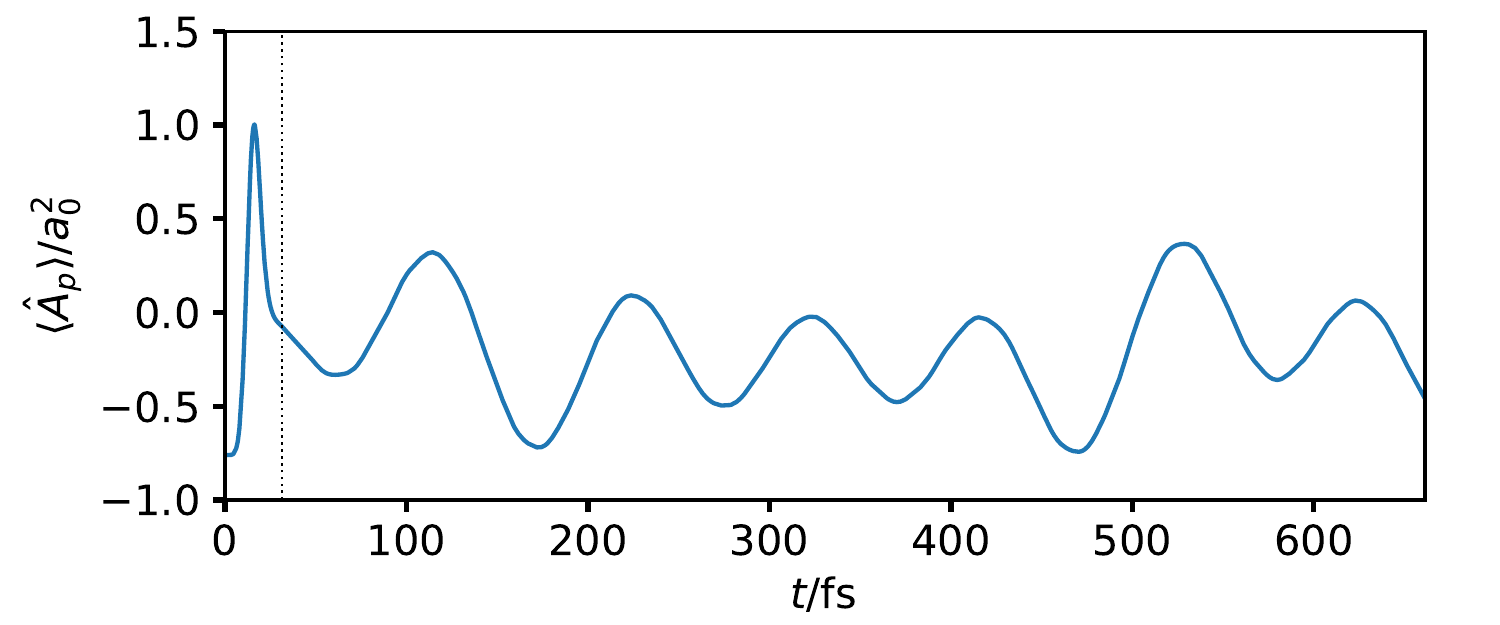}
\caption{\label{fig:Ap}$\braket{\hat{A}_p}$ during and after interaction
with the $100$-cycle ($31.5\,\text{fs}$) resonant laser pulse. The vertical dotted line marks the end of the laser pulse.
}
\end{figure}

The expectation value $\braket{\hat{A}_p}$ is plotted in Fig.~\ref{fig:Ap}.
In qualitative agreement with BO theory,~\cite{Stapelfeldt2003}
the highest degree of alignment is observed slightly after the laser intensity peaks (at $t=16.25\,\text{fs}$)
and approximately
periodic revivals of the maxima are observed after the pulse is turned off. The period is roughly $102\,\text{fs}$ for both
maxima and minima.

\begin{figure*}
    \centering
     \begin{subfigure}[b]{0.48\textwidth}
         \centering
         \includegraphics[width=\textwidth]{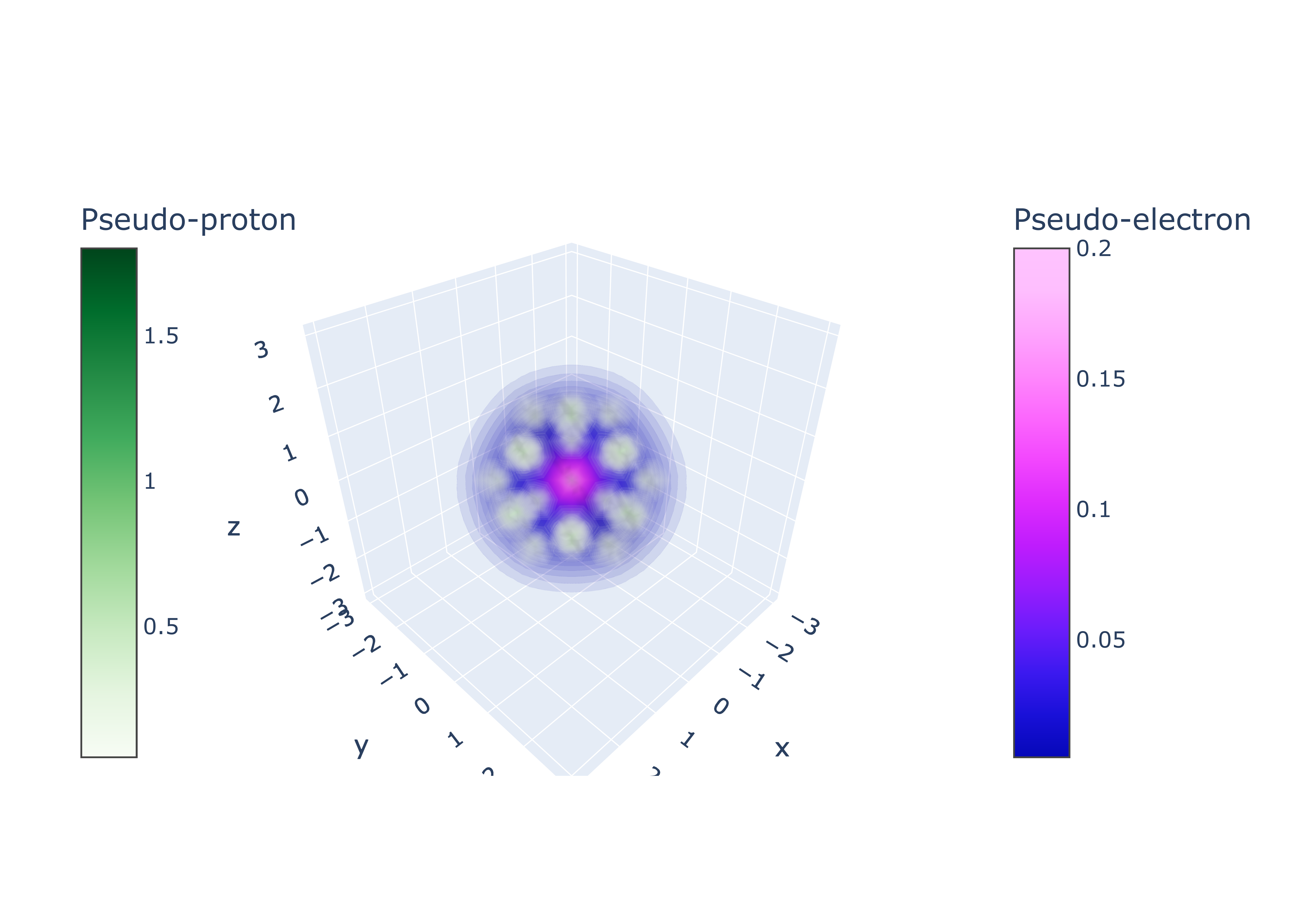}
         \caption{$t = 0.00\,\text{fs}$, $\braket{\hat{A}_p} = -0.7611$}
         \label{fig:Ap_initial}
     \end{subfigure}
     \hfill
     \begin{subfigure}[b]{0.48\textwidth}
         \centering
         \includegraphics[width=\textwidth]{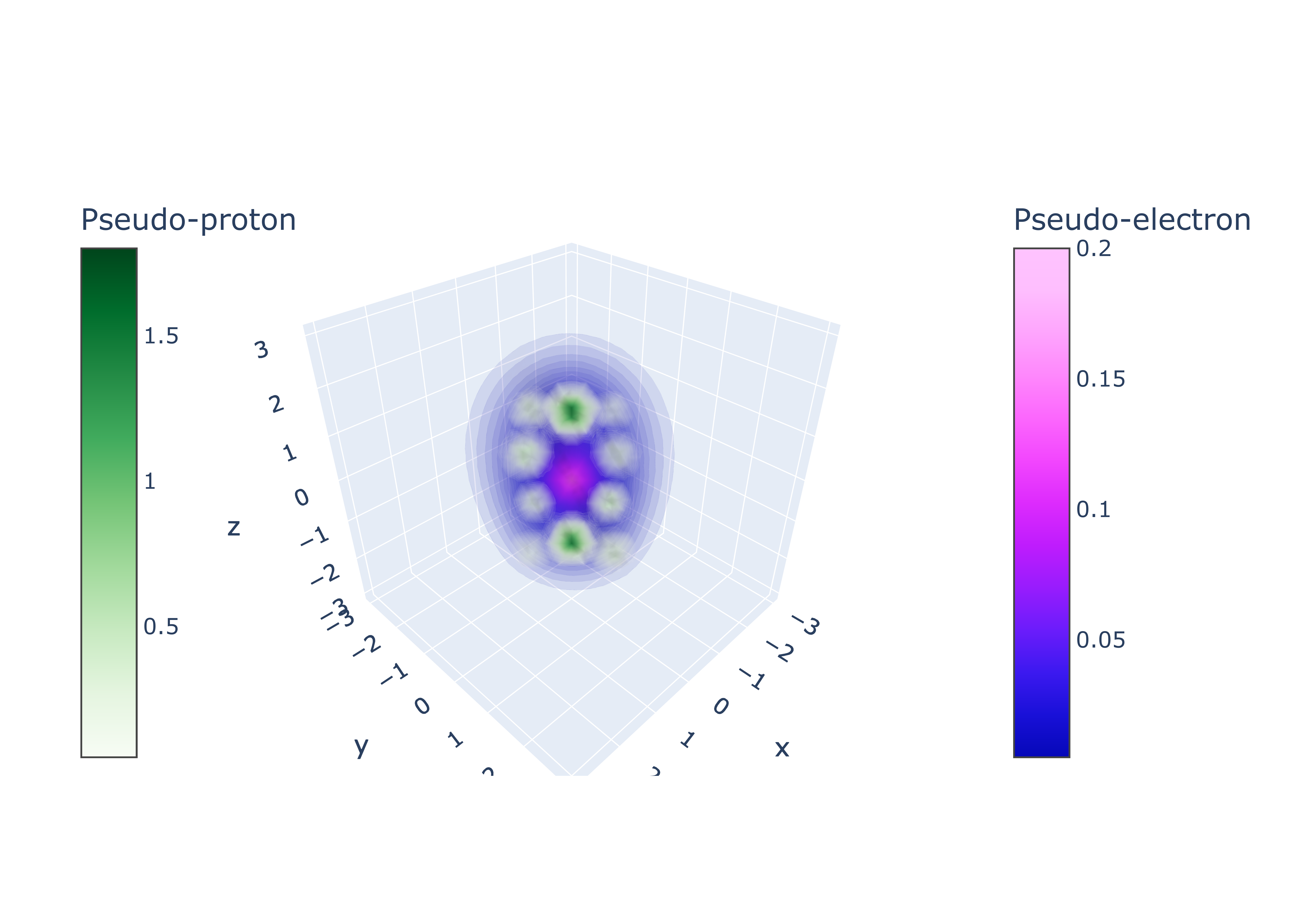}
         \caption{$t = 16.25\,\text{fs}$, $\braket{\hat{A}_p} = 1.0026$}
         \label{fig:Ap_max0}
     \end{subfigure}
     \newline
     \centering
     \begin{subfigure}[b]{0.48\textwidth}
         \centering
         \includegraphics[width=\textwidth]{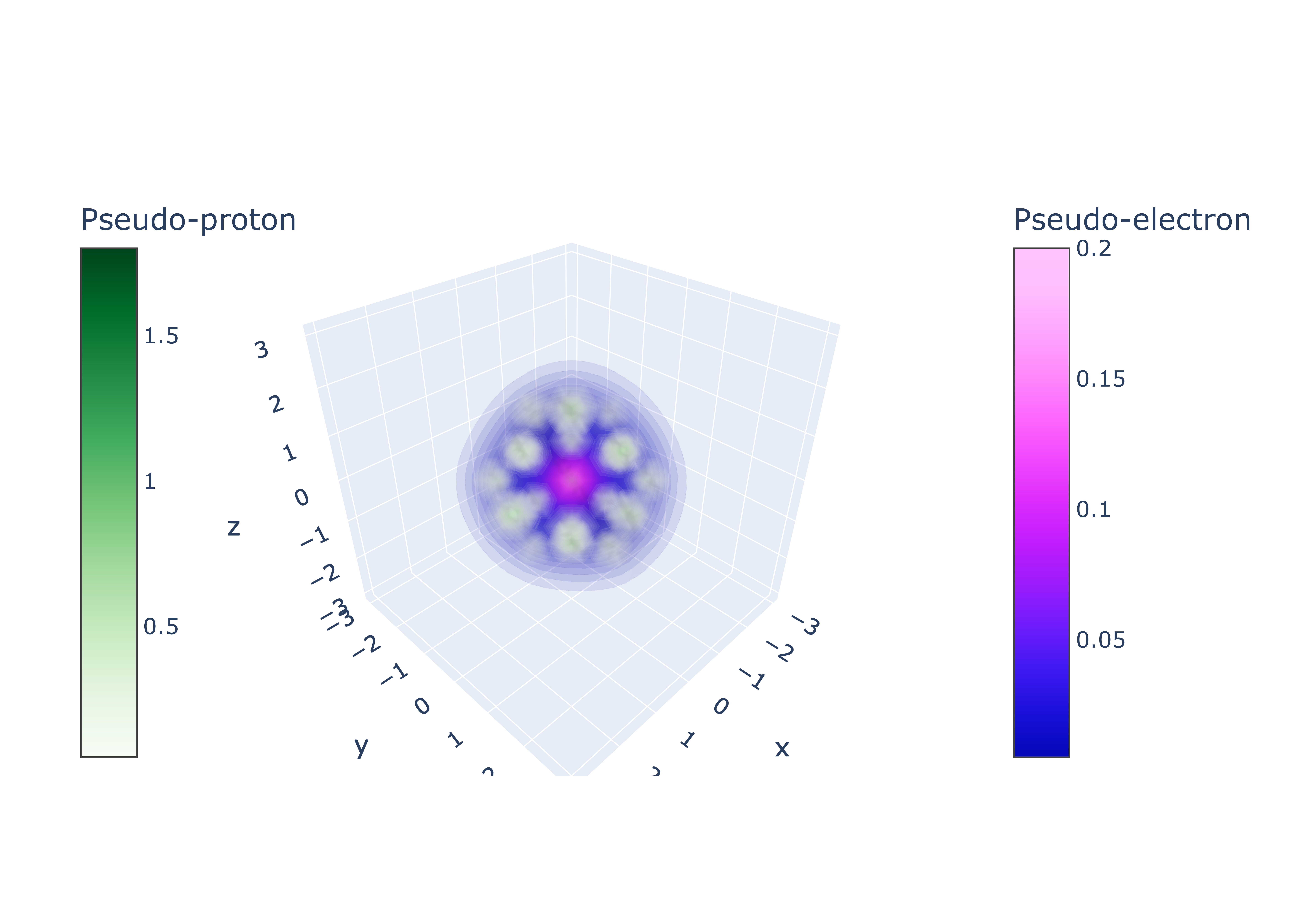}
         \caption{$t = 469.94\,\text{fs}$, $\braket{\hat{A}_p} = -0.7430$}
         \label{fig:Ap_min4}
     \end{subfigure}
     \hfill
     \begin{subfigure}[b]{0.48\textwidth}
         \centering
         \includegraphics[width=\textwidth]{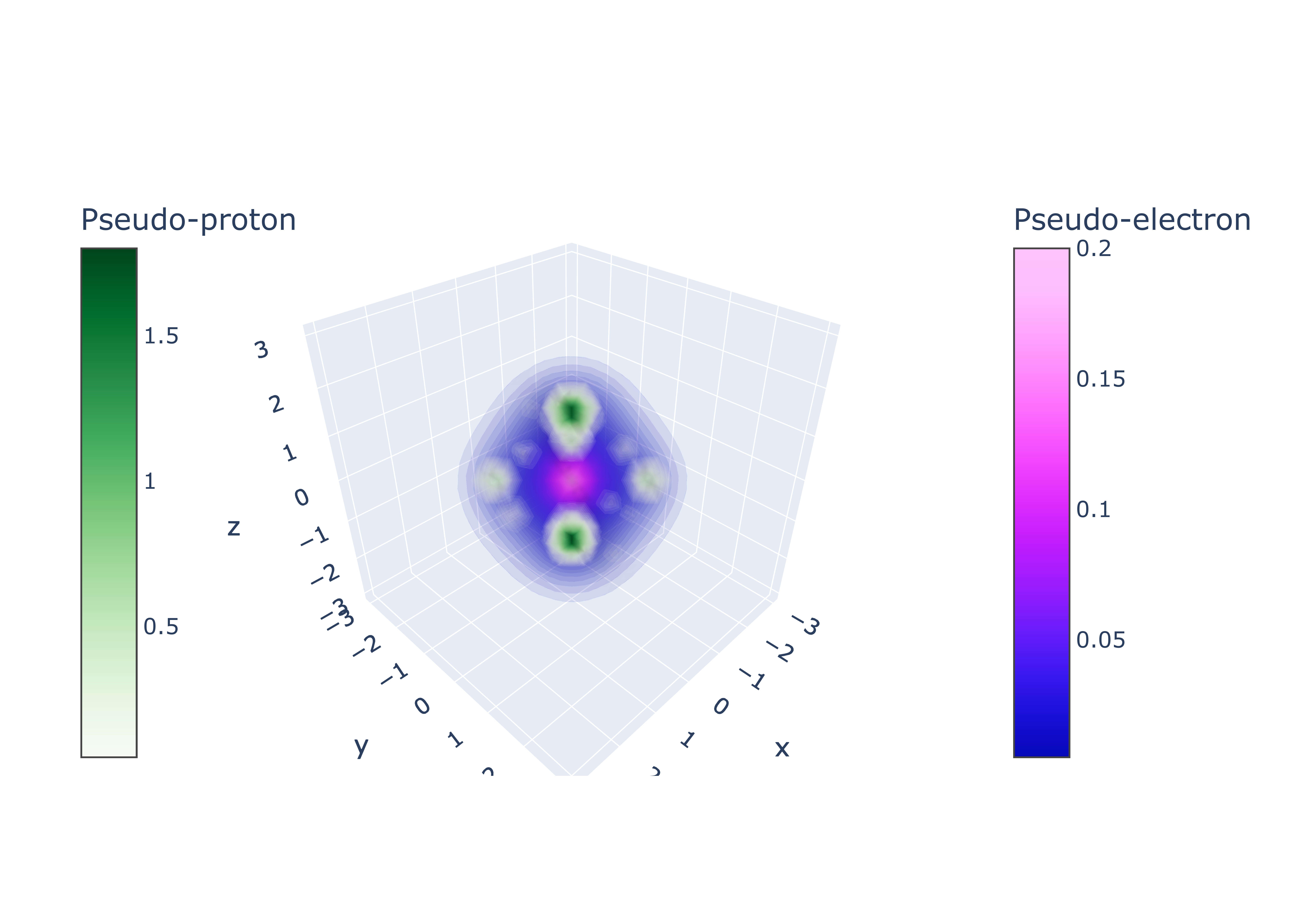}
         \caption{$t = 528.67\,\text{fs}$, $\braket{\hat{A}_p} = 0.3663$}
         \label{fig:Ap_max5}
     \end{subfigure}
     \caption{\label{fig:Ap_snapshots}HD pseudo-particle densities during and after interaction with the $100$-cycle pulse.
     }
\end{figure*}

The alignment can be visualized through the pseudo-proton density, which is shown along with the pseudo-electron density
in Fig.~\ref{fig:Ap_snapshots} at selected time steps.
The pseudo-particle densities are computed from the $48$ dominant stationary states
of the internal molecular wave packet (reproducing the norm of the wave function to within $0.5\%$ throughout the whole dynamics)
using a slight generalization of the approach described in Ref.~\onlinecite{Cafiero2005}.
Although spherical symmetry is not enforced \emph{a priori}, the basis-set construction procedure yields essentially
correct symmetry of the ground-state densities, Fig.~\ref{fig:Ap_initial}. Slightly after the laser intensity peaks,
the pseudo-proton density peaks sharply on the $z$-axis, corresponding to alignment. The off-axis density appears
to be a non-BO manifestation of the BO concept of pendular states.~\cite{Friedrich1991,Friedrich1995,Friedrich1995a}
After the pulse is turned off, the densities oscillate between ground-state-like spherical distributions, Fig.~\ref{fig:Ap_min4},
and predominantly aligned distributions, Fig.~\ref{fig:Ap_max5}. Although the pseudo-proton density peaks even more
sharply than at the peak laser intensity, the value of $\braket{\hat{A}_p}$ is significantly reduced due to spreading
in the $xy$-plane. These perpendicular components arise during the dynamics in the second half of the pulse.

\begin{figure}
\centering
\includegraphics[width=\columnwidth]{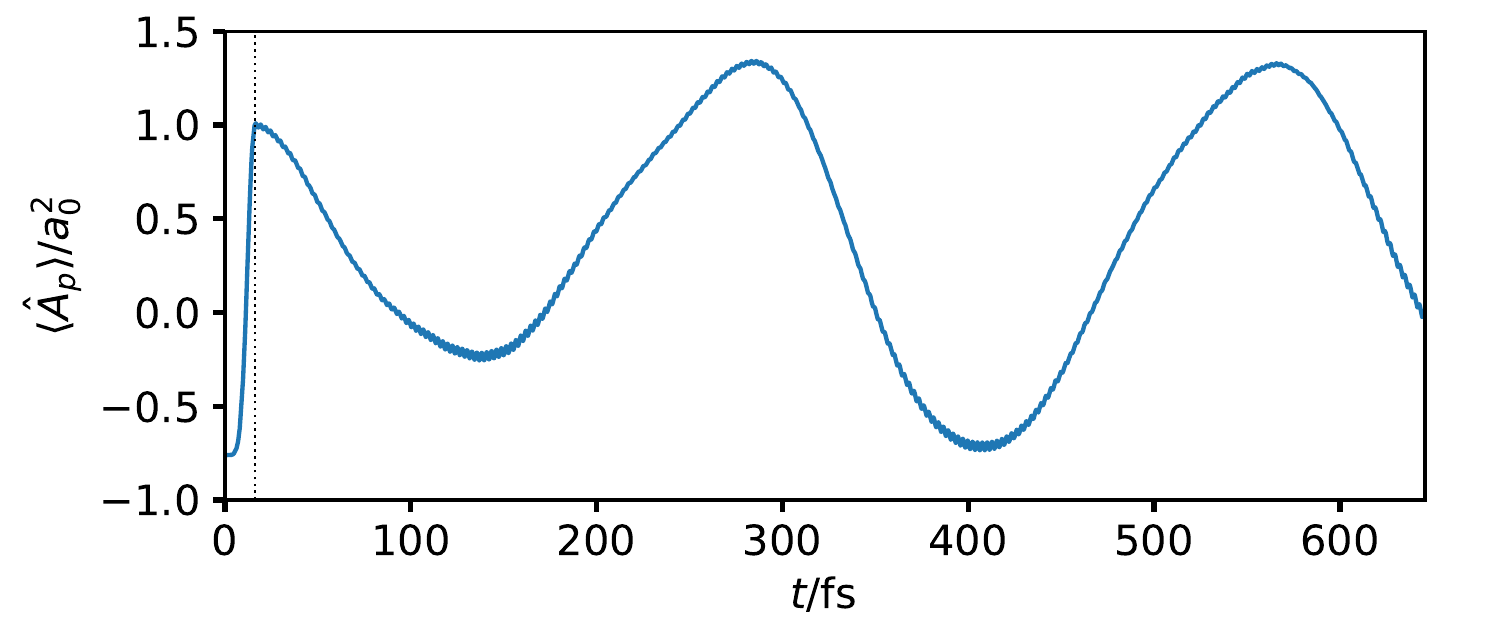}
\caption{\label{fig:Ap2}$\braket{\hat{A}_p}$ during and after interaction
with the ($50+1$)-cycle ($16.1\,\text{fs}$) resonant laser pulse. The vertical dotted line marks the end of the laser pulse.
}
\end{figure}

\begin{figure*}
    \centering
     \begin{subfigure}[b]{0.48\textwidth}
         \centering
         \includegraphics[width=\textwidth]{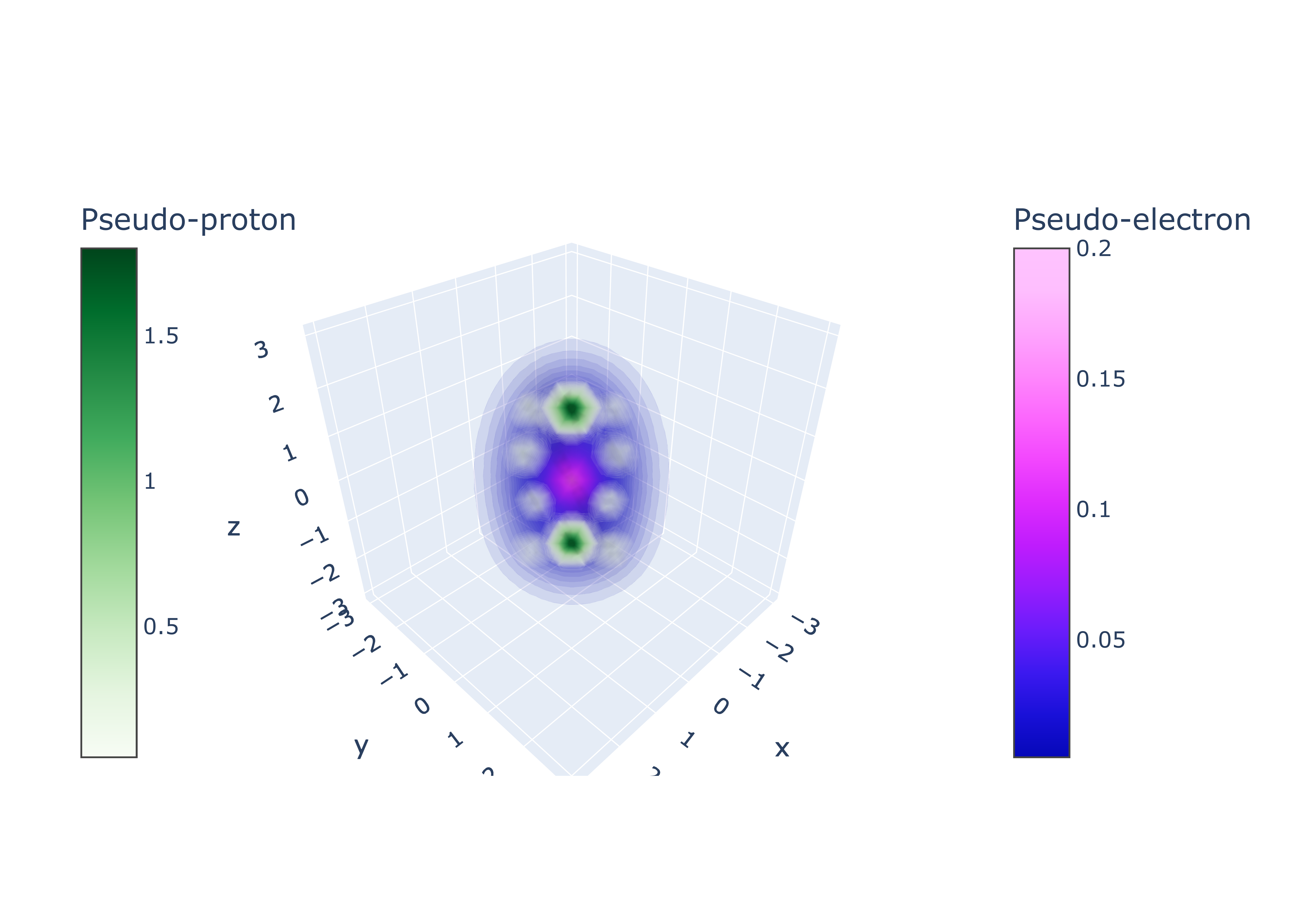}
         \caption{$t = 16.55\,\text{fs}$, $\braket{\hat{A}_p} = 1.0087$}
     \end{subfigure}
     \hfill
     \begin{subfigure}[b]{0.48\textwidth}
         \centering
         \includegraphics[width=\textwidth]{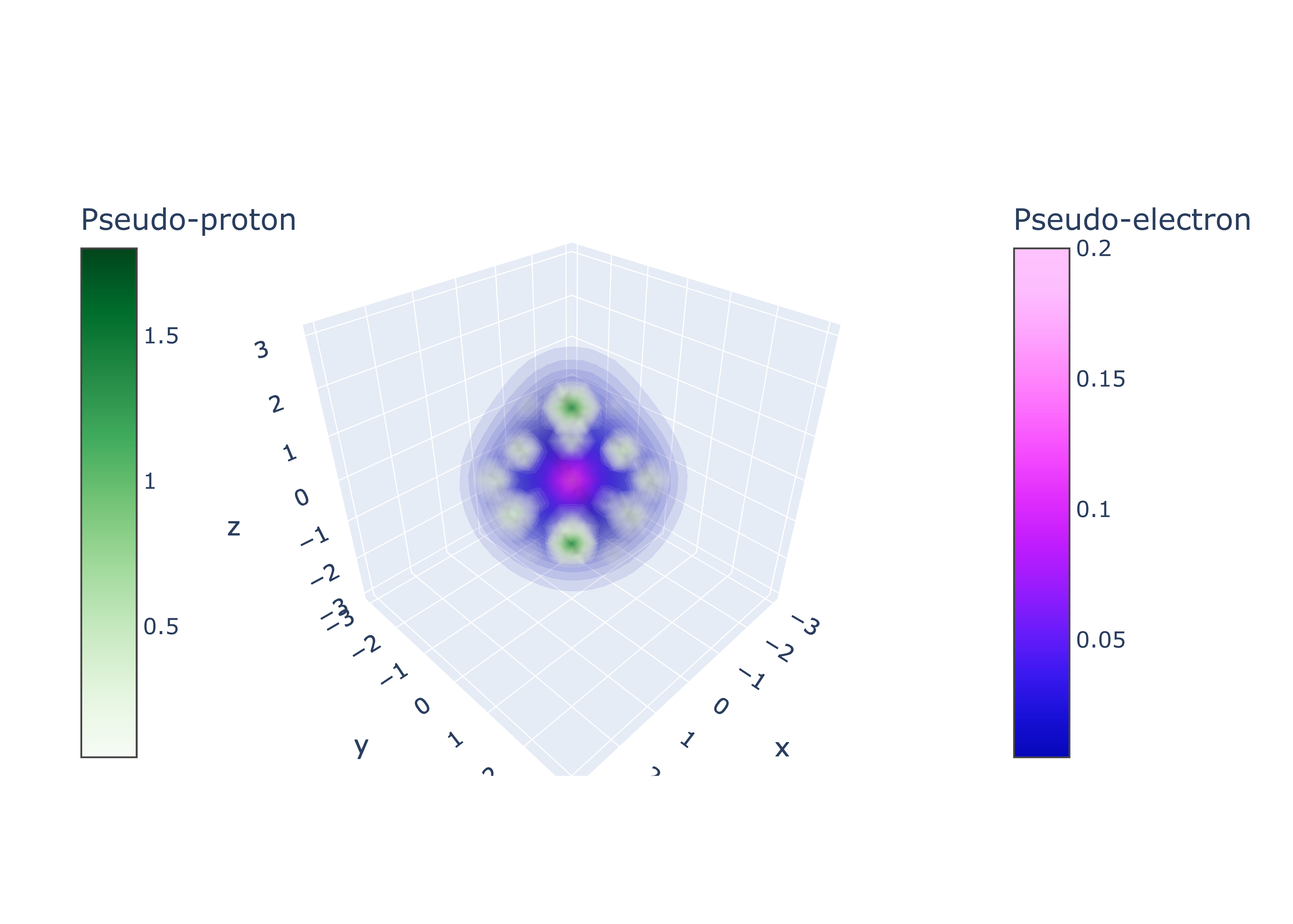}
         \caption{$t = 136.93\,\text{fs}$, $\braket{\hat{A}_p} = -0.2555$}
     \end{subfigure}
     \newline
     \centering
     \begin{subfigure}[b]{0.48\textwidth}
         \centering
         \includegraphics[width=\textwidth]{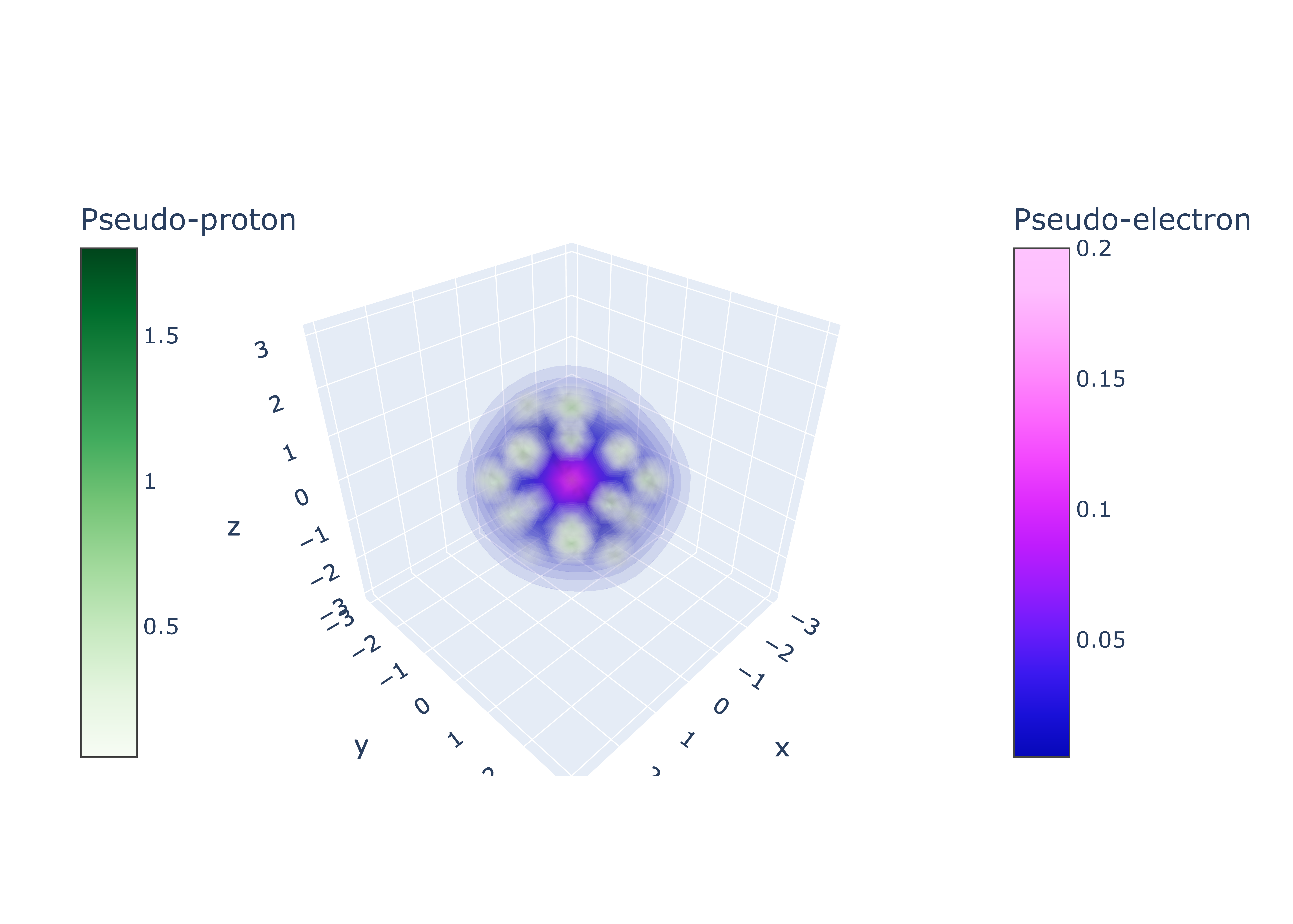}
         \caption{$t = 406.23\,\text{fs}$, $\braket{\hat{A}_p} = -0.7346$}
     \end{subfigure}
     \hfill
     \begin{subfigure}[b]{0.48\textwidth}
         \centering
         \includegraphics[width=\textwidth]{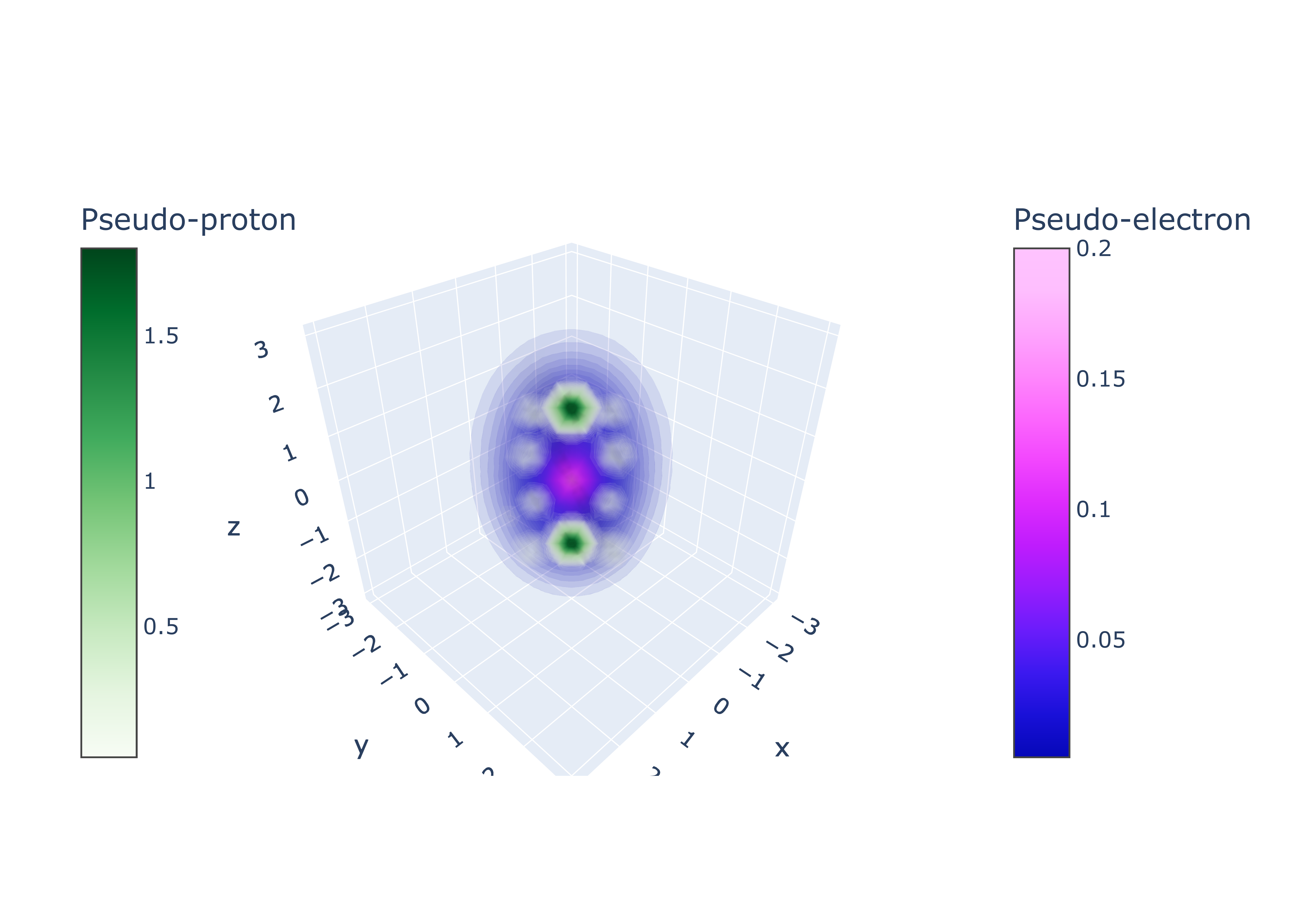}
         \caption{$t = 565.63\,\text{fs}$, $\braket{\hat{A}_p} = 1.3295$}
     \end{subfigure}
     \caption{\label{fig:Ap2_snapshots}HD pseudo-particle densities during and after interaction with the ($50+1$)-cycle pulse.
     }
\end{figure*}

Inspired by the ``adiabatic turn-on, rapid turn-off'' approach,~\cite{Yan1999}
we perform an additional simulation where the same laser pulse is very rapidly turned off after the peak
intensity is reached at $50$ cycles. Specifically, we use a $1$-cycle turn-off time. The resulting $\braket{\hat{A}_p}$
is shown in Fig.~\ref{fig:Ap2}.
With the ($50+1$)-cycle pulse, the oscillation period is significantly longer: the average time between maxima
and between minima is roughly $275\,\text{fs}$. The $\braket{\hat{A}_p}$ curve indicates enhanced alignment
during the post-pulse dynamics, an observation confirmed by the density plots in Fig.~\ref{fig:Ap2_snapshots}.

\section{Concluding remarks}

Laser-induced molecular alignment is well established experimentally and well understood within the theoretical framework
of the BO approximation. In this communication,
we have presented proof-of-principle non-BO simulations
of the HD molecule exposed to resonant laser pulses, showing the emergence of field-free alignment in the sense of
the pseudo-proton density oscillating between essentially spherical and linear shapes.
We have used a limited number of static ECG basis functions built from variational bound-state optimization
in the presence of static electric fields.
Our results thus are qualitative, and
quantitative simulations
would require more flexible ECG basis sets constructed with time-dependent complex nonlinear parameters,
enabling high-accuracy non-BO simulations of, e.g., high-harmonic generation spectra from aligned few-particle molecules.

\begin{acknowledgments}
The authors acknowledge the support of the Centre for Advanced Study in Oslo, Norway, which funded and hosted our CAS research project
\emph{Attosecond Quantum Dynamics Beyond the Born-Oppenheimer Approximation}
during the academic year 2021/2022.
The work was supported by the Research Council of Norway through its Centres of Excellence scheme, Project No.\ 262695,
and simulations were performed on resources provided by Sigma2 - the National Infrastructure for High Performance Computing and 
Data Storage in Norway (Project No.\ NN4654K).
Partial support from the National Science Foundation (grant No. 1856702)
is also acknowledged.
\end{acknowledgments}

\nocite{*}
\bibliography{final}

\end{document}